\documentstyle[12pt,epsf]{article}
\title{Re-examination of  long distance effects  in $b \to s l^+ l^- $ }
\author{ Cai-Dian  L\"{u} and Da-Xin Zhang\\
 Physics Department, Technion- Israel Institute of Technology, \\
  Haifa 32000, Israel.}

\date{}
\begin{document}
\maketitle

\begin{abstract}
We re-analyse the long distance contributions
to the process $b \to s l^+ l^-$. 
Full $q^2$-behavior of the vector meson dominance amplitude
is used together with the effect of Terasaki suppression,
and comparisons with the previous results are given.
We show that the interference between short- and long distance contributions
makes it difficult to extract the short distance information 
from the dominant long distance background,
 either in the dilepton
invariant mass distribution or in the single lepton energy spectrum.
\end{abstract}
PACS number(s): 13.40.Hq, 12.15.Ji, 13.20.Jf
\\
\newpage

Rare  decays  through the flavor changing $b\to s$ 
transitions provide good test of the standard model (SM), and 
are expected to give signals of new physics \cite{smrev}. 
The branching ratio of the process $b\to s\gamma$, 
which has already been measured by the 
CLEO collaboration\cite{bkstar}, 
is within the SM predictions. 
Unlike the decay  $b\to s\gamma$,
the process $b\to s l^+l^-$($l=e$ or $\mu$)  is expected to be dominated by long distance 
contributions through the mechanism of vector meson dominance(VMD)\cite{res}. 
However, it was  usually believed  that the long 
distance (resonance) contributions  arise only in some particular region 
of the invariant mass spectrum of the dilepton pair\cite{bsll}, 
since the involved resonance $\psi (\psi ')$ peak is very sharp. 
Detailed calculation\cite{res} shows that there 
exists significant interference between the short and long distance 
contributions, 
which leaves only a small portion of kinematic region
at low dilepton invariant mass
where the  interference effect by the resonances  is small.  
The energy spectrum of single lepton has also been given in \cite{don}  
where a  window of nearly pure short distance information is found.

In the present work, we will re-examine both the dilepton invariant mass 
and the single lepton
 energy spectrums using an alternative treatment of the VMD amplitude.
In the previous analysis of the cascade decays
$b\to s\psi(\psi')$ and $\psi(\psi')\to l^+l^-$,
an effective description  is made for the later electromagnetic transition 
$\psi(\psi')\to l^+l^-$\cite{res,don}. 
The dependence of the  VMD amplitude on 
the square of the dilepton invariant mass, $q^2$,
is approximated by that of the resonance mass $m_{\psi}^2$ or $m_{\psi'}^2$
in the denominator of the photon propagator.
This approximation is only valid near the resonance region,
and consequently,  the previous  
analysis are not complete in the whole phase space\footnote{This 
has also been noted in \cite{wise} in the discussion of
exclusive decays of B meson, and in
\cite{singer} in the charm quark sector.}.
 
Let us start with the short distance contributions to $b \to s l^+l^-$
with $l=e$ or $\mu$.
The short distance contributions come from  box, $Z$ and photon penguin 
diagrams. The QCD corrected effective Hamiltonian in SM is \cite{bsll}:
\begin{equation}
{\cal H}_{eff} =\frac{\alpha G_F }{\sqrt{2}\pi}
 V_{tb}V_{ts}^* \left[ \left (C_9^{eff} \bar s
\gamma^\mu P_L b +\frac{2C_7^{eff} m_b}{q^2} \bar s \not \! q \gamma^\mu 
P_R b\right ) \bar l\gamma_\mu l \; + C_{10} (\bar s\gamma^\mu P_L b)
\bar l\gamma_\mu \gamma_5 l\right],\label{1}
\end{equation}   
with $P_L=(1-\gamma_5)/2$, $P_R=(1+\gamma_5)/2$, and $q=p_{l^+} +p_{l^-}$ is 
the invariant mass of the dilepton. 
The analytic Wilson coefficients $C_7^{eff}(\mu)$, $C_9^{eff}(\mu)$, and $C_{10}(\mu)$ are 
given in Ref.\cite{ali}.
Under the leading logarithmic approximation, we get the numerical results 
at $\mu=m_b$ as:
\begin{equation}
C_7^{eff}=-0.315, ~~~C_{10}=-4.642,
\end{equation}
and to the next-to-leading order,
\begin{eqnarray}
C_9^{eff}&=& 4.227+0.124~ \omega (\hat s )
+0.359~ g(m_c/m_b,\hat s) +0.034~ g(1,\hat s) +0.033~g(0,\hat s) ,
\end{eqnarray}
where $\hat s=q^2/m_b^2$.
The function  $\omega (\hat s )
$ and $g(z,\hat s)$ can be found in ref.\cite{ali}.
Here for numerical evaluation, 
we use $m_{top}=176$GeV\cite{cdf}, $m_b=4.8$GeV,
$m_c=1.4$GeV, $\Lambda_{QCD}=225$MeV
\cite{pdg}.

By normalizing to the semileptonic rate, the strong
dependence on the b-quark mass cancels out.
The differential decay rate $d\Gamma(
\overline{B} \rightarrow X_s l^+l^-)/d \hat s$, where $\hat s =( p_{l^+}
+p_{l^-})^2/m_B^2$, is given by
\begin{eqnarray}
\frac{1}{\Gamma (\overline{B} \rightarrow X_c e \overline{\nu})}
 \frac{d}{d\hat s} \Gamma(\overline{B} \rightarrow X_s l^+l^-)
&=& \frac{\alpha_{QED}^2}{4\pi^2
 f (m_c/m_b)} (1-\hat s)^2 \left[ (1+2\hat s) 
 \left( |C_9^{eff}|^2 +C_{10}^2\right)\right .\nonumber\\
&+& \left .4\left( 1+\frac{2}{\hat s} \right) |C_7^{eff}|^2 +12 C_7
 Re(C_9^{eff}) \right ],\label{3}
\end{eqnarray}
where  $f(m_c/m_b)$ is the phase space factor:
$$f(x)=1-8x^2+8x^6-x^8-24x^4\ln x.$$
If we take  the experimental result $Br(\overline{B} \to
X_c e\overline{\nu} ) =10.8\% $\cite{pdg}, 
the differential decay rate  of
$\overline{B} \to X_s \mu^+\mu^-$ is found, which is depicted in Fig.1 as 
the dash-dotted line.

In addition, there are also long distance resonance contributions 
from $c\bar c$ state.
There are six known resonances in the $c\bar c$ system that can contribute 
to this decay mode \cite{ref}. 
The lowest two, $\psi$ and $\psi'$ ,
were considered in the previous analyses\cite{res,don}. 
Here  we 
also consider the same two resonances. 
The higher resonances will also contribute, 
but they are less important in our case of discussing
the uncertainties,
as will be shown later. 

Applying the VMD mechanism, 
the long distance contribution is through $b \to s \psi$, 
and $\psi\to \gamma \to l^+l^-$, 
where the resonance can also be $\psi'$.
These give the effective Lagrangian 
\begin{equation}
{\cal L}_{res} = \frac{16\pi^2 a_2 g_\psi^2(q^2) }{ 3q^2 (q^2 -m_\psi^2 
+im_\psi\Gamma_\psi)} \; \frac{\alpha G_F}{\sqrt{2}\pi} V_{tb}V_{ts}^*  
(\bar s\gamma^\mu P_L b) \bar l\gamma_\mu  l ~+~(\psi\to \psi '),\label{4}
\end{equation}
where $a_2=C_1 +C_2/3$ is a QCD corrected  coefficient of the four quark 
operators. 
Below we will use the phenomenological value $a_2=0.24$
which comes from  fitting  the data of $B$ meson decays\cite{phen}. 
Note that the expression (\ref{4}) 
differs from the previous ones\cite{res,don} by keeping the photon 
propagator as $-ig^{\mu\nu}/q^2$ instead of $-ig^{\mu\nu}/m_\psi^2$ or $-ig^{\mu\nu}/m_{\psi'}^2$.
Thus it holds in the whole kinematic region.

The effective coupling of a vector meson $g_V(q^2) (V=\psi\;,\;\psi' \;)$ is 
defined by
\begin{equation}
<0|\bar c\gamma_\mu c|V (q)>=g_V(q^2)\epsilon_\mu^V,
\end{equation}
where $\epsilon_\mu^V$ is the polarization vector of the vector meson $V$.
On the mass-shell of the vector meson,
$g_V(q^2)$ is replaced by
the decay constant $g_V(m^2_V )$ which can be obtained from the leptonic 
width of the vector meson:
\begin{equation}
\Gamma (V\to\ell^+\ell^- )=\frac{16\pi\alpha^2}{27m_V^3}g_V^2(m^2_V).
\label{6}
\end{equation}

The structure of
eqn.(\ref{4})  is  the same as that of the operator $O_9$. 
It is convenient to include the resonance contribution in eqn.(3) by simply 
making the replacement  
\begin{equation}
C_9^{eff}\to C_9^{\prime eff}=C_9^{eff}+ \frac{16\pi^2 a_2 g_\psi^2(q^2) }{ 3q^2 (q^2 -m_\psi^2 
+im_\psi\Gamma_\psi)}\; +\;(\psi\to \psi ').
\end{equation}
Assuming a constant coupling $g_\psi^2(q^2)\equiv g_\psi^2(m_\psi^2)$
as done in \cite{res,don},
the numerical result is given in Fig.1 as the dashed line. 
It is easy to see that,
this spectrum is enhanced  in the low $q^2$ region due to the explicit
 inclusion of the photon propagator. 
From  Fig.1, we can also expect that higher resonances other than $\psi$ or $\psi'$
 contribute mainly in the region  $0.6<\hat s<1$ 
where we are not interested,  
since near this tail of the spectrum no useful short distance
information is expected to emerge.

The assumption made above on the constant coupling of $g_\psi^2(q^2)$
can be improved by accounting for the mechanism  of Terasaki suppression
for the  $\psi -\gamma$ conversion\cite{tt}.
In the framework of VMD, 
the data on photoproduction of $\psi$ indicates a large
suppression of $g_\psi (0)$ compare to $g_\psi (m^2_\psi )$\cite{dht}.
This has been confirmed in \cite{eims} by constraining the dominant 
long distance contribution to $s\to d\gamma$ using the present upper
bound on the $\Omega^-\to\Xi^-\gamma$ decay rate.  
As a result, it can be concluded
that this  suppression results in a much smaller long distance
contribution to $b\to s\gamma$ transition\cite{dht}. 
Now we use a momentum 
dependent $g_V(q^2)(V=\psi\; ,\; \psi' )$ in ${\cal L}_{\rm res}$, 
which is used in  \cite{am} to obtain a reduced resonance to
nonresonance interference where a broader region of invariant
mass spectrum sensitive to short distance physics is claimed.

The momentum dependence of $g_V(q^2)$($V=\psi\; ,\;\psi'$)  derived 
 using a dispersion relation\cite{tt} is
\begin{equation}
g_V(q^2)=g_V(0)\left (1+\frac{q^2}{c_V}\left [d_V -h(q^2)\right ]\right ),
\label{8}
\end{equation}
where $c_\psi =0.54\; ,\; c_{\psi'}=0.77$ and $d_\psi =d_{\psi'}=0.043$.
$h(q^2)$ is defined by
\begin{equation}
\displaystyle
h(q^2)=\frac{1}{16\pi^2r}\left \{ -4-\frac{20r}{3}+4(1+2r) 
\sqrt{\frac{1}{r}-1} ~~{\rm tan}^{-1}\frac{1}{\sqrt{\frac{1}{r}-1}}\right \}
\end{equation}
with $r=q^2/m_V^2$ for $0\leq q^2\leq m_V^2$.  
As a result, eqn.(\ref{8}), 
which is valid for $0\leq q^2\leq m_V^2$, 
is an interpolation of $g_V$ from the photoproduction
experimental data on $g_V(0)$ to $g_V(m_V^2)$ from the
leptonic width based on quark-loop diagram.  
We assume $g_V(q^2)=g_V(m_V^2)$ for $q^2 > m_V^2$ mainly due
to the fact that the behavior of the $\psi -\gamma$ conversion strength is
not clear in this region, and is not important in our case (see below).  

Applying  Terasaki's formula (\ref{8}) for  the
$q^2$ dependence of $g_V(q^2)$, 
the differential decay rate of $b\to s l^+l^-$ receives suppression in low 
$q^2$ region. However, there is still significant interference between the 
resonance  and the short distance contributions, 
due to the factor $1/q^2$ coming from the propagator of the virtual photon. 
This is also shown in Fig.1.

Now we turn to the energy spectrum  of single lepton. 
The integration over $q^2$ is 
complicated, since many functions here involve $\hat s$. 
We simply give the numerical results in Fig.2 
for $l=\mu$(see also \cite{don}). 
One can see that,
if the $1/q^2$ behavior is replaced by $1/m_\psi ^2$ (or $1/m_{\psi'} ^2$) everywhere,
there is almost no contribution from the $\psi$, $\psi'$ resonance when 
$\beta < 0.2$. 
This result is  what has been arrived in Ref.\cite{don}. 
If, however, this $1/q^2$ is retained, 
there are also contributions from the resonances even in the low $\beta$ region
and consequently,
the resonance background is still serious. 
Including  the effect of Terasaki suppression  of $g_V(q^2)$ in the low 
$q^2$ region, the result is also shown in Fig.2 where 
the resonance background is only half  reduced.

 From both Fig.1 and Fig.2, 
we can observe that the  long distance VMD contributions
to the process $b\to sl^+l^-$ are large, 
if an alternative treatment of the electromagnetic sub-process is performed.
It can also be seen that the single lepton energy spectrum is
almost useless in extraction of any short distance information
from the resonance background. 
The total branching ratio of $b\to s l^+l^-$ turn out to be  
$3.6\times 10^{-4}$ or  
 $4.9\times 10^{-4}$  with  
the effect of Terasaki suppression  included or not.

We have treated the resonance contribution from $\psi$, $\psi'$
to $b\to sl^+l^-$ alternatively
without using the effective description of the electromagnetic
sub-process $\psi(\psi')\to \gamma\l^+l^-$.
The long distance contributions are found to be significant,
especially in the  single lepton energy spectrum. 
Concerning other higher resonances, 
and also other uncertainties existed in  this decay mode\cite{wise},
we conclude that it is 
difficult to extract  short distance information, which is sensitive to
new physics, from the dominant long 
distance contributions.

\section*{Acknowledgement}

We thank G. Eilam, M. Gronau and P. Singer for helpful discussions.
The research of D.X. Zhang is supported in part by Grant 5421-3-96
from the Ministry of Science and the Arts of Israel.

\vskip 0.9in
\section*{Figure Captions}

\vskip 0.2in

\noindent
Fig.1 The differential decay rate via $\hat s =m_{\mu\mu}^2 /m_b^2$. 
The  dash-dotted and the dotted lines correspond to the results without
resonance contribution and  
with resonance contribution included as \cite{res,don}, respectively.
The  dashed and solid lines are the results
with resonance included with the  treatment of the 
electromagnetic  sub-processes using eqn. (5),
while constant coupling $g_V(q^2)=g_V(m_V^2)$ and (\ref{8})
are used, respectively.

\vskip 0.2in

\noindent
Fig.2  Same as Fig.1 for the differential decay rate via   $\beta =E^- /m_b$.

\newpage
\begin{figure}
\centerline{\epsffile{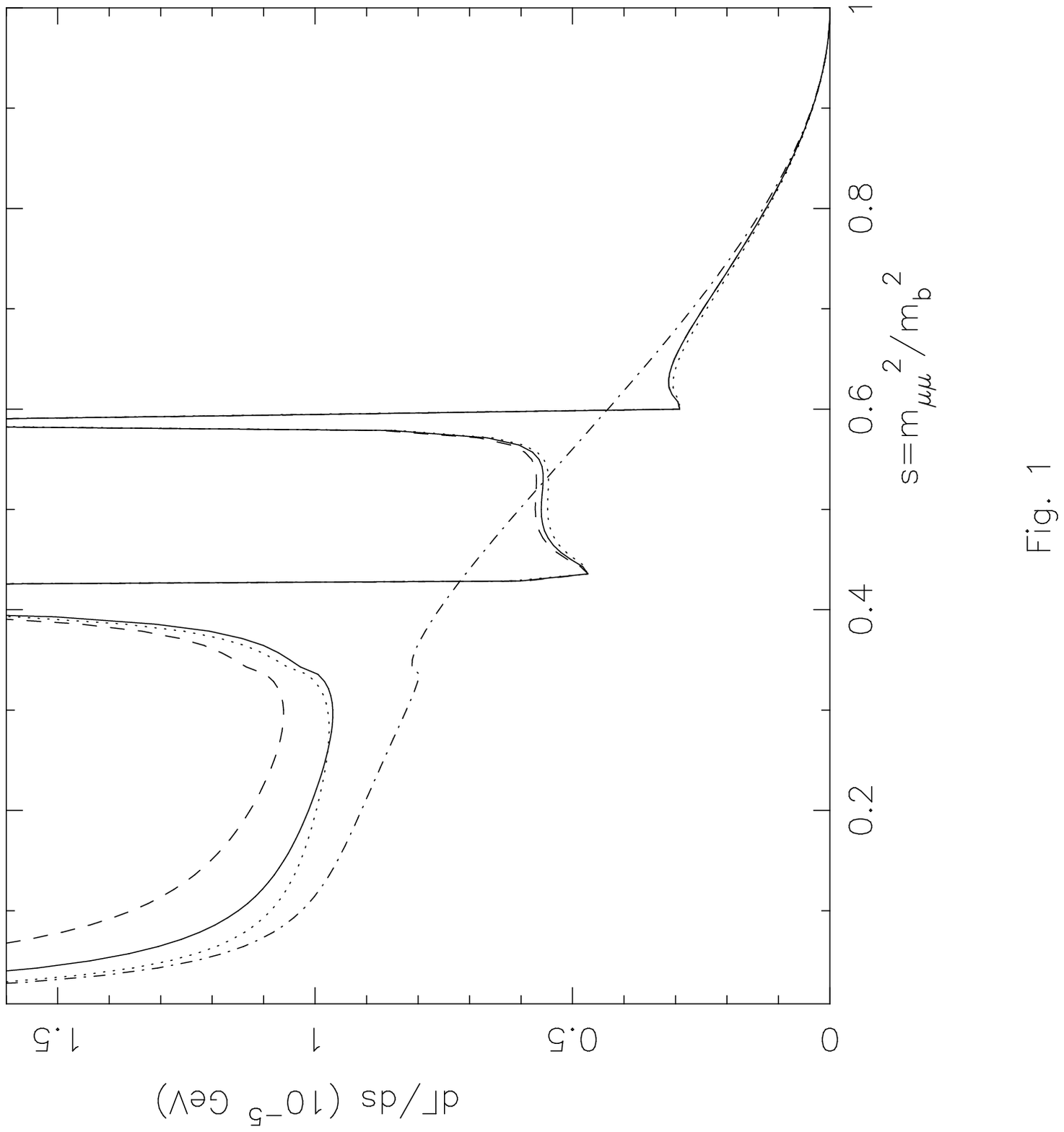}}
\end{figure}

\begin{figure}
\centerline{\epsffile{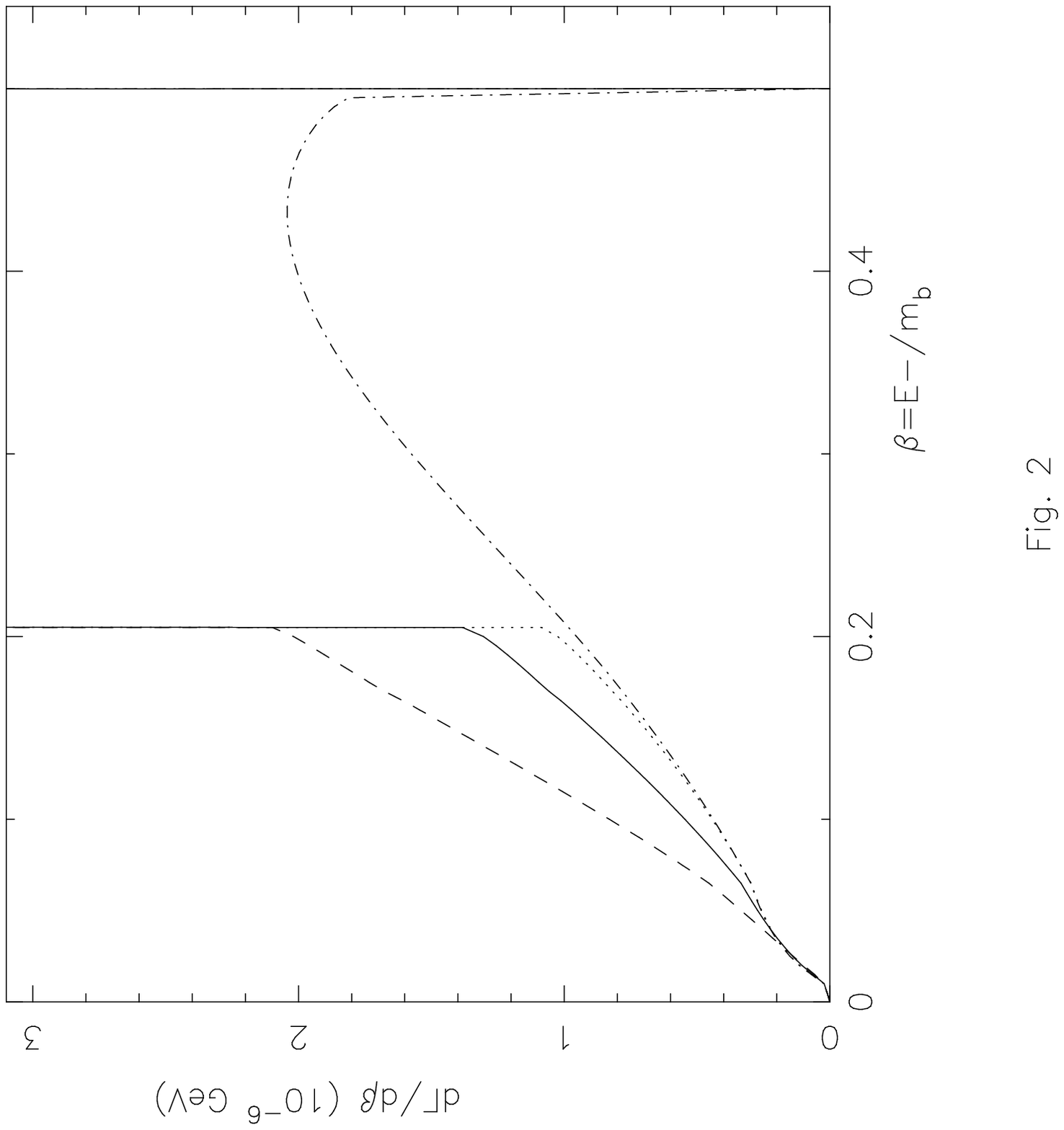}}
\end{figure}

\end{document}